# A USER-FRIENDLY APPROACH TO LOCAL ANODIC OXIDATION

Pasqualantonio Pingue[*]
*Laboratorio NEST-INFM, Scuola Normale Superiore, Piazza dei Cavalieri 7, 56126 Pisa, Italy*

Paolo Baschieri, Cesare Ascoli
*Istituto per i Processi-Chimico Fisici del CNR (IPCF – CNR), Area della Ricerca di Pisa, Via G. Moruzzi 1, 56124 Pisa, Italy*

Michel Dayez
*Centre de Recherche de la Matiere Condensé e et des Nanoscience  (CRMC-CNRS), Campus de Luminy, 13288 Marseille Cedex 9, France*

## Abstract

A home made DSP-controlled scanning probe microscope (SPM) system has been developed and its utilization as a tool for lithography on nanometer scale is reported. User-friendly graphic interface allows to directly perform nanolithography importing a bitmap pattern on previously-imaged sample region, otherwise moving in real time the probe using an ordinary mouse, acting as a pantograph from macroscopic down to nanometer scale, taking advantage from an absolute-positioning stage. Any measured property of the sample can be used by the lithographic interface to realize interactive patterning of the sample. Moreover, the "*contour-mode*" permits to move tip at a fixed relative tip-sample distance, exploiting at each step the previously-acquired topographic signal. The instrument has been successfully employed in order to perform local anodic oxidation (LAO) lithography on semiconducting substrates.

---

[*] Electronic address:  pingue@sns.it

**Introduction**

In the last few years a growing interest has been devoted to nanolithography due to integrated circuit critical dimension shrinking. For this reason more demands have been requested on lithographic techniques. Among these, certainly scanning probe lithography (SPL) [1] satisfies those demands for high resolution achievable in different configurations and for its intrinsic auto-alignment capabilities. Ploughing [2], local anodic oxidation (LAO) [3], and dip pen nanolithography (DPN) [4] are the most important and promising SPL techniques.

In this paper the "NanoWorker", an integrated DSP-controller/SPM-head system for lithography and mapping at nanometer scale, is presented and its applications to LAO nanolithography are reported.

**Experimental**

*NanoWorker* is basically an AFM system designed to make easy its use as nanotool. It can work as an AFM for imaging in three different configurations (contact mode, intermittent-contact mode and acquiring a whole set of force-distance curves) and alternatively it can be used to modify the sample.

The key point for the *NanoWorker's* user is the graphic interface where the acquired image is represented. This interface makes it possible moving the tip directly by mouse in real time or to transfer a bitmap image on the sample by lithography: the program will operate the transfer acting as a pantograph from macroscopic down to nanometer scale. The method to be used for the lithographic transfer is previously selected from the control panel. High-speed data handling of *NanoWorker* allows using the graphic interface on line, in real time. It is possible to move the probe on the sample surface towards selected points by using an ordinary mouse, performing approach or withdraw by a simple mouse's click and changing lithographic parameters as force or voltage using the mouse scroll. In the imaging mode *NanoWorker* can acquire up to 5 different physical quantities so that the criteria to select the points to be modified employing the lithographic mode can be based on several different characteristics of the material. In particular, when force-distance curve imaging is employed, selection can be based for example also on spectroscopic data, like adhesion or elasticity maps.

The experimental setup is based on an home-built AFM-head equipped with a commercially-available sample stage (120 µm scan range in x and y axis, 15 µm in z axis), having absolute positioning capability in x-y (noise ~ 2 nm rms) and z (~ 1 nm rms) axis trough capacitive sensors and its relative closed-loop feedback [6]. In Fig.1 there is a schematic of our system. The chassis of the AFM-head is made in Invar in order to minimize thermal drift on nanometer scale. The DSP-based controller, a completely stand alone modular unit, is home built and we can install inside how many A/D and D/A converter cards we need. We can disconnect computer and reconnect it at any time during the experiment: the system provide an automatic resynchronisation. This feature ensures that scanning and feedback stability are not affected in case of OS crash. PC Software has been developed in Labwindows-CVI environment.

A commercially available modulation and AC-to-rms conversion unit [7] was employed for intermittent-contact mode (tapping mode) and phase imaging in air and liquid, switching to standard contact-mode imaging for nanolithography and nanomanipulation. Commercially available $W_2C$-coated and n-doped silicon conductive tips on V-shaped cantilevers [8] were employed both for standard imaging as for the other processes.

The software is constituted by four main windows: 1) the "parameter window"; 2) the "data acquisition window"; 3) the "vectorial window"; 4) the "bmp window". The first one allows setting the *NanoWorker* operating mode and all the related parameters. The other three windows are those that allow performing mapping, lithography and nanomanipulation.

**Results**

*NanoWorker* has been successfully employed to perform imaging in three different configurations: i) contact mode; ii) intermittent-contact mode; iii) "Touch & Lift" [9] mode. These configurations are managed by the data-acquisition window that allows visualizing in real time the acquired channel (also as a 3D view) or to elaborate the acquired image *a posteriori*. More in detail, *Touch & Lift* scanning method allows to acquire force-distance curves on each point of the scanned area so to obtain (with some data processing when necessary) force, stiffness and adhesion maps of the sample. For example, it can be also employed for current-distance maps when the AFM is configured in conductive mode (as for example in the case of meniscus-conductance studies in LAO process [10]).

The developed software permits a true lithographic capability in various configurations. When the topographic image is acquired (in contact, tapping mode or *Touch & Lift* configuration) and the interesting region is localized on the sample, it is possible to graphically import the acquired image in the "vectorial window" (see Fig.2). In the so called "*manual-litho*" mode, keeping the left button down, the AFM tip can be moved through each pixel of the image, following in real time the mouse movements inside this window. It is also possible to assign the starting and the ending points of a linear trajectory using two separate left button clicks. Tip approach or withdraw to the sample surface is ensured by one click of the right button (approach is performed employing the *Touch & Lift* approach procedure). Finally, pushing and rotating the mouse scroll we can change tip-sample force or voltage values. Each process is performed in real time and is possible to monitor interesting parameters (topography, lateral force, tip-sample bias or current etc.) in the five-channel flow chart. When imaging is performed in tapping mode or *Touch & Lift* configuration, the system automatically approach the tip and switches to constant force mode (without any hardware or software change) to perform lithography. Nevertheless, in "*automatic-litho*" mode the pattern is previously drawn by the user in the lithography window, and then followed by the tip automatically at the set velocity, force or voltage. The same process can be also stored and repeated, since the system keeps all the parameters and the same timing previously employed. Moreover, during the LAO lithography a voltage modulation can be employed, removing in this way the charge build-up on the produced oxide[11]. Both the modulation amplitude as the frequency can independently be changed and different waveforms can be chosen (squared, sinousodal, DC) with a tunable voltage offset. Force-distance curve in few selected points can be utilized in order to check the presence of water meniscus between the sample and the tip so that the necessary external humidity conditions for LAO are verified. Fig. 2 reports the final oxide pattern obtained on *n*-doped GaAs by vectorial lithography LAO.

In this case a DC bias in output was employed to drive in current the anodization process employing a voltage-current converter (1 µA/1V ratio). Whereas Fig. 3 shows how to perform LAO in automatic mode on GaAs/AlGaAs heterostructure, depleting in this way the two-dimensional electron gas underneath in order to fabricate, as in this case, a quantum dot nanostructure [12, 13]. The whole pattern was "hand made" employing the *NanoWorker* in manual-litho configuration using the computer mouse pointer as a pen: the oxide lines are approximately 50-nm large and 5-nm thick. Referring to Fig.3, blue lines are region where the tip is up (withdrawn) and the bias zero, whereas red lines are the trajectories where tip is down in contact, the system in force-feedback and the bias is applied to the sample. As explained before, approach procedure down to the force set point is the same that is normally employed for force-distance maps. Notice the alignment capability of the system that permits adding new features after imaging on a previously patterned structure. Low-voltage SEM picture of the final structure is also reported, where detailed structure of the produced oxide stripes is clearly shown without any tip artifact, and where electrical discharges appear as small spherical spots along the stripes.

In the vectorial window we also implemented a new lithographic feature called "*contour mode*" that allows moving the tip in real time (by mouse) on the surface at a fixed relative tip-sample distance, exploiting at any point the previously-acquired topographic signal. In this configuration, it is possible to upload the sample topography, setting the distance (positive or negative) from the sample surface and using this parameter as reference to keep the tip-sample distance constant. The tip moves (in real time) following in vectorial mode the sample surface contour above/under the surface itself at the set-point value of the distance. A left mouse button defines the path to follow by point and click operation, right-button click switches between *contact mode* and *contour mode*, and the mouse scroll can be employed to change the applied force, the applied voltage or the relative distance in real time during the lithographic process. A straightforward application of *contour mode* is LAO in non-contact, exploiting a water meniscus "stretching" and therefore improving the final resolution [14]. An example of this lithographic technique is reported in Fig. 4-a,b where double-tip writing effect has been removed by meniscus stretching. Tilting of the sample surface was present along the diagonal axis of the image (~10º) and compensated by the *contour mode*. Direct reading of the z-axis movement can be performed and acquired in real time on the flow chart during the lithographic process in order to monitor the *contour* process. Notice also silicon tip wearing at the end of the LAO process, as seen in low-voltage SEM pictures, that evidence a "volcano" aspect probably related to the electrochemical process associated with the oxidation itself (Fig. 4-c).

Finally, as additional feature, *NanoWorker* can work in "*smart BMP-mode*" trough a specific control panel. This panel allows to superimpose a bitmap image on a selected region of the acquired signal and to perform lithography or mapping just in that region. To perform "BMP-litho" we have to select the required pattern uploading it as inset in a previously-imaged sample region as a standard 24 bit RGB bitmap. *Smart BMP-mode* configuration, inset dimension and positioning are chosen by the operator in the "acquisition-data window" employing the mouse pointer, and the relative BMP-window pops-up automatically (see Fig.4-1). As previously shown, the lithography or mapping is performed just when the tip passes on the inset during the standard rastering. This peculiar configuration improves the quality of the lithographic process also in the case where no absolute positioning is available, due to the fact that a new tip-positioning is not necessary, so there are not added hysteresis and creep phenomena to the scan. Moreover, the lithographic function is performed just in forward direction while backward acquisition permits to have imaging in real time, line-by-line, of the modified surface. This feature results very useful, for example, in order to check immediately if the lithographic process is going on correctly and to avoid any further unintentional modification during imaging on very soft substrates or patterns.

In the simplest BMP-litho configuration, it is possible to associate a particular lithographic function when the tip is on a particular colour tone of the bitmap itself. For example, it is possible to perform LAO associating to different RGB values a different tip-sample bias (see Fig.5-(a),(b)) or to growth oxide only in selected regions, where a particular colour is present. In this configuration the AFM behaves like a "LAO plotter", wrinting line by line the final structure (see Fig. 5-(c),(d)). Also in this case, low-voltage SEM imaging allows highlighting the single oxide stripes in the letter "&", removing any tip artifact.

"*Smart BMP-mode*" allows also performing logic operations between up to three different bitmaps (related to different acquired channels or to a previously-defined pattern) and, using an appropriate selection of colors trough the integrated graphic tools, it is possible to emphasize a particular physical property of the sample as topography, friction or elasticity. This particular feature permits to accomplish a "smart" lithography or mapping that acts differently according to local physical properties of the sample, acting for example as "topography (spectroscopy)-sensitive litho". This capability can be employed in fact in order to acquire, only in the interesting regions of the sample, force-distance or current-distance maps, avoiding acquisition-time losses. Fig. 6 shows a simulation of "*smart-bitmap mode*" process on a tapping-mode imaged biological sample (a mould in this case). The resulting pattern can be employed to perform selectively mapping or lithography just on the interesting region. Finally, "*Smart BMP-mode*" can be employed as a tool for an interactive alignment markers finder in multi-step lithographic processes.

**Conclusions**

A new SPM-based system for lithography, the *NanoWorker*, has been presented and its application to local anodic oxidation has been demonstrated. New lithographic modalities, *smart-BMP* and *contour*, have been introduced and few interesting applications reported. Low-voltage scanning electron microscopy has been also proved as an excellent analysis tool for LAO structures characterization.

# References

[1] For a review see: *"Technology of Proximal Probe Lithography"*, edited by C. R. K. Marrian, SPIE, Bellingham, WA, **1993**, and references therein.

[2] (a) T.A. Jung , A. Moser, H.J. Hug, D.Brodbeck, R. Hofer, H.R. Hidber and U.D. Schwarz, *Ultramicroscopy* **1992,** 42-44, 1446; (b) R. Magno and B.R. Bennett, *Appl. Phys. Lett.* **1997**, 70, 1855; (c) P. Pingue, M. Lazzarino, F. Beltram, C. Cecconi, P. Baschieri, C. Frediani, and C. Ascoli, *J. Vac. Sci. Technol. B* **1997,** 15, 1398.

[3] J. A. Dagata, J. Schneir, H. H. Harary, C. J. Evans, M. T. Postek, and J.Bennett, *Appl. Phys. Lett.* **1990,** 56, 2001.

[4] Richard D. Piner, Jin Zhu, Feng Xu, Seunghun Hong, Chad A. Mirkin, *Science* **1999,** 283, 661.

[5] T. Junno, K. Deppert, L. Montelius, and L. Samuelson, *Appl. Phys. Lett.* **1995,** 66, 3627.

[6] *Queensgate*, UK. Internet site address: http://ww.nanopositioning.com.

[7] *Elbatech* srl, Italy. Internet site address: http://www.elbatech.com.

[8] *NT-MDT*, RU. Internet site address: http://www.ntmdt.ru.

[9] Cappella B., Baschieri P., Frediani C., Miccoli P. and Ascoli C., *Nanotechnology* **1997,** 8, 82.

[10] F. Perez-Murano, C. Martin, N. Berniol,, H. Kuramochi, H. Yokoyama and J.A. Dagata, *Appl. Phys. Lett.* **2003**, 82, 3086.

[11] (a) R. Garcia, M. Calleja, and P. Perez-Murano, *Appl. Phys. Lett.* **1998,** 72, 2295; (b) B. Legrand and D. Stievenard, *Appl. Phys. Lett.* **1999**, 74, 4049.

[12] S. Luscher, A. Furher, R. Held, T. Heinzel, K. Ensslin, and W. Wegscheider, *Appl. Phys. Lett.* **1999,** 75, 2452.

[13] T. Heinzel, S. Luescher, A. Fuhrer, G. Salis, R. Held, K. Ensslin, W. Wegscheider, and M. Bichler, *Proc. SPIE Int. Soc. Opt. Eng.* **2000,** 4098, 52.

[14] R. Garcıa, M. Calleja, and H. Rohrer, *J. Appl. Phys.* **1999,** 86, 1898.

**FIGURES**

**Figure 1** Schematic of the *Nanoworker* system. The two-buttons&scroll mouse plays an essential role in lithography (see text).

**Figure 2** The vectorial window (see text). Upper panels: drawing "Italia" by LAO on a previously-patterned *n*-doped GaAs substrate. Bottom panel: final result.(scan size: 4 μm).

**Figure 3** LAO performed in automatic mode on GaAs/AlGaAs heterostructure. (a) Vectorial window with the defined pattern on mesa structure. (b) Final LAO result by low-voltage SEM imaging.

**Figure 4** (a) Schematic between LAO in contact mode and in *contour mode*. (b) LAO on GaAs at different distances (tip/sample distance: 15, 25 and 35 nm), employing the same tip-sample bias (+10 V). Scan range: 7.2 μm. (c) SEM images of a silicon tip before and after the LAO process.

**Figure 5** (a) "*Smart BMP-window*" with the inserted bitmap file. Green&Blue colorscale corresponds to 0-100 pA range. (b) Corresponding SEM image of the LAO pattern on GaAs obtained in *smart-BMP* configuration. (c) AFM image of a BMP logo written by LAO on GaAs. (d) Detail of "&" letter imaged by low-voltage SEM.

**Figure 6** "*Smart-BMP mode*" applied on a biological sample (a mould). Clockwise: uploading the bmp file; positioning it in the "BMP-window" on the acquired image; low-threshold color settings; high-threshold color settings; AND logic operation; NOT logic operation. The last green region can be employed for LAO on the semiconductor substrate.

**FIGURE 1**

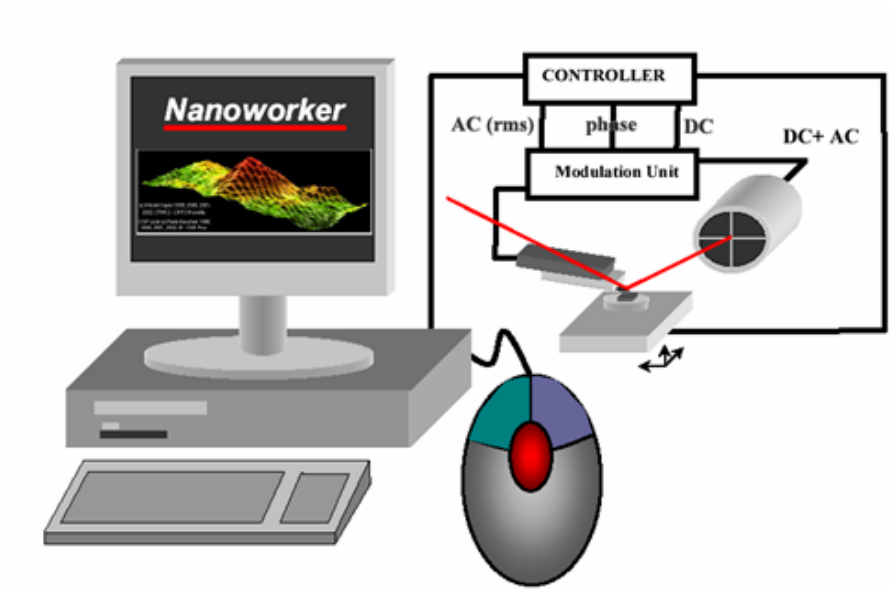

P. Pingue *et al.*

**FIGURE 2**

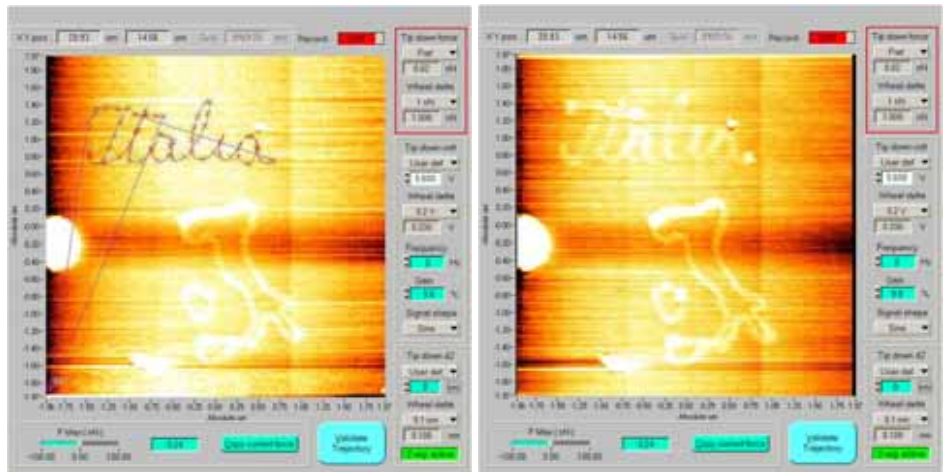

P. Pingue *et al.*

**FIGURE 3**

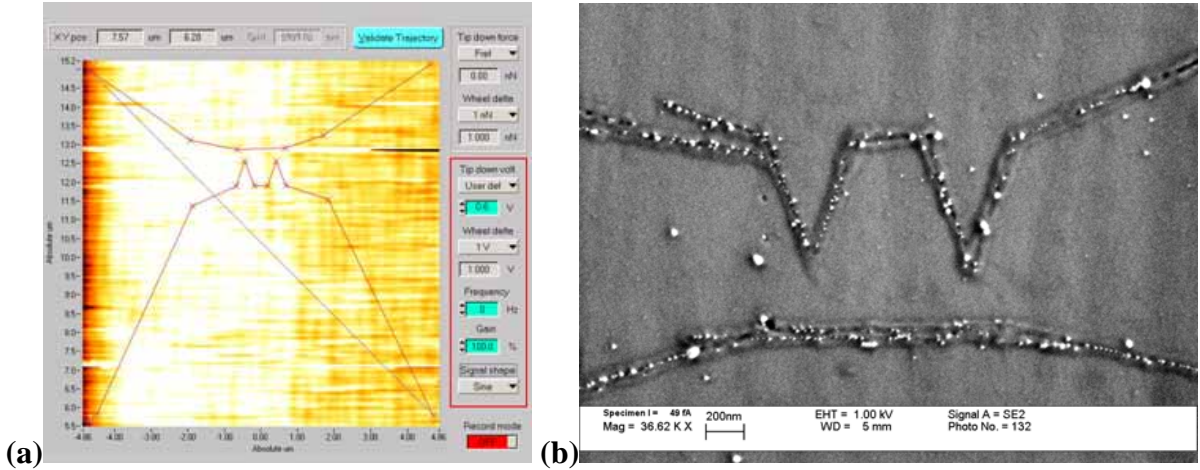

(a) (b)

P. Pingue *et al.*

**FIGURE 4**

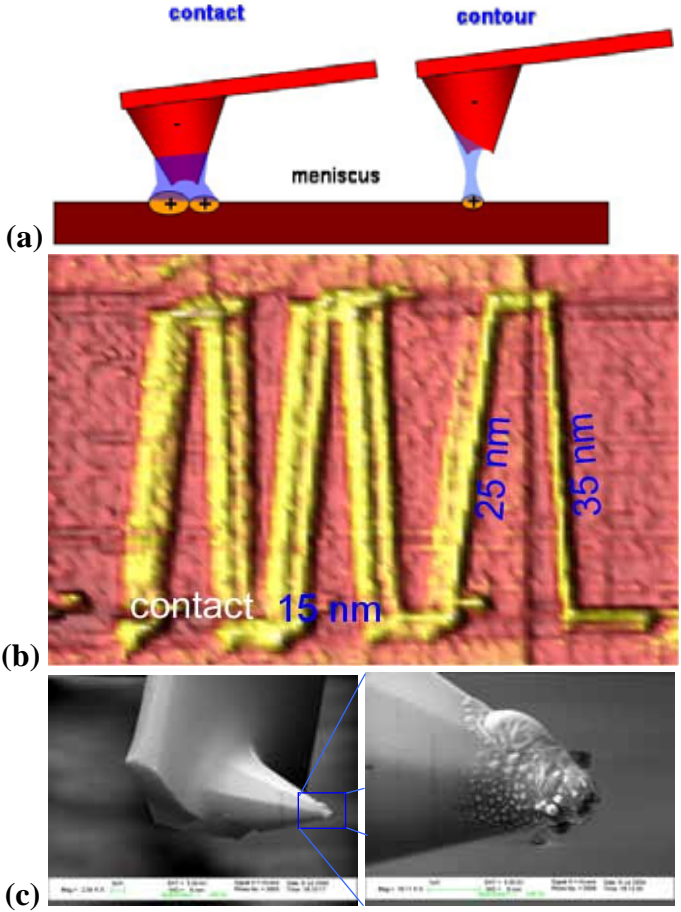

P. Pingue *et al.*

**FIGURE 5**

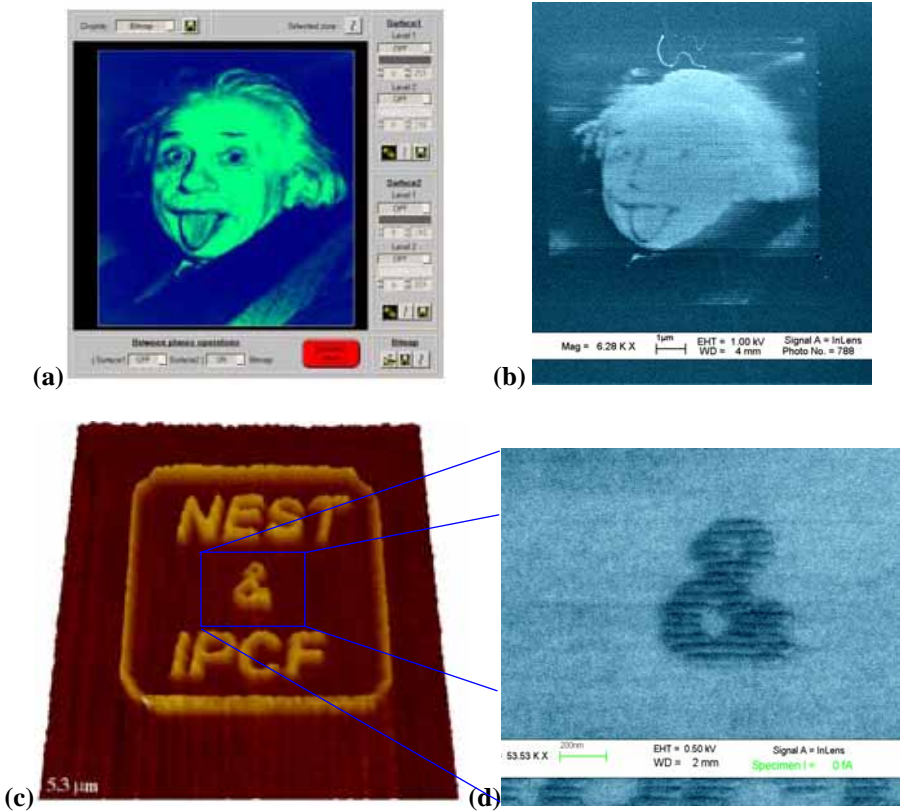

(a) (b) (c) (d)

P. Pingue *et al.*

**FIGURE 6**

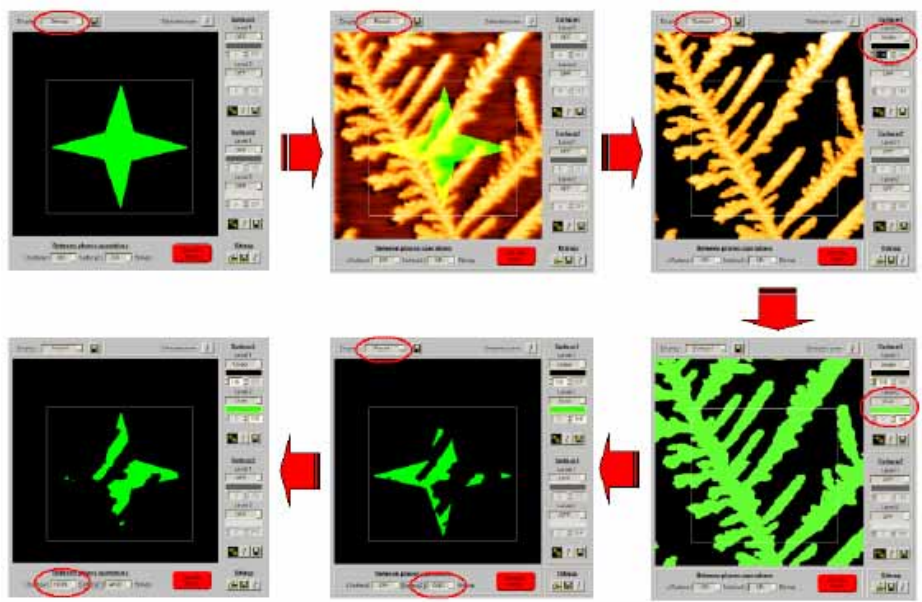

P. Pingue *et al.*